\documentclass[%
 reprint,
 amsmath,amssymb,
 aps,
]{revtex4-2}

\usepackage{graphicx}% Include figure files
\usepackage{dcolumn}% Align table columns on decimal point
\usepackage{bm}% bold math
\begin{document}

\title{About carrier's self-trapping and dynamical Rashba splitting in the two-dimensional hybrid perovskite (BA)$_2$(MA)$_2$Pb$_3$I$_{10}$}

\author{W. Qi$^{1}$, S. Ponzoni$^{1}$, G. Huitric$^{1}$, V. Gorelov$^{1}$, A. Pramanik$^{1}$, Y. Laplace$^{1}$, M. Marsi$^{2}$, E. Papalazarou$^{2}$, S. F. Maehrlein$^{3,4,5}$, E. Deleporte$^{6}$, N. Mallik$^{7}$, A. Taleb‐Ibrahimi $^{7}$, A. Bendounan $^{7}$, K. Zheng$^{8,9}$, T. Pullerits$^{9}$ and L. Perfetti$^{1}$}
\affiliation
{$^{1}$ Laboratoire des Solides Irradi\'{e}s, CEA/DRF/lRAMIS, Ecole Polytechnique, CNRS, Institut Polytechnique de Paris, F-91128 Palaiseau, France}
\affiliation
{$^{2}$ Universit\'e Paris-Saclay, CNRS, Laboratoire de Physique des Solides, 91405 Orsay Cedex, France}
\affiliation
{$^{3}$ Department of Physical Chemistry, Fritz Haber Institute of the Max Planck Society, 14195, Berlin, Germany}
\affiliation
{$^{4}$ Institute of Radiation Physics, Helmholtz-Zentrum Dresden-Rossendorf, 01328, Dresden, Germany}
\affiliation
{$^{5}$ Institute of Applied Physics, Dresden University of Technology, 01187, Dresden, Germany}
\affiliation
{$^{6}$ Universit\'e Paris-Saclay, ENS Paris-Saclay, CentraleSup\'elec, CNRS, Lumi\`ere, Mati\`ere et Interfaces (LuMIn) Laboratory, 91190 Gif-sur-Yvette, France}
\affiliation
{$^{7}$ Synchrotron SOLEIL, Saint Aubin BP 48, Gif-sur-Yvette F-91192, France}
\affiliation
{$^{8}$ Department of Chemistry, Technical University of Denmark, DK-2800 Kongens Lyngby, Denmark}
\affiliation
{$^{9}$ Chemical Physics and NanoLund, Lund University, Box 124, 22100 Lund, Sweden}

\begin{abstract}

Time- and Angle-Resolved Photoelectron Spectroscopy (tr-ARPES) is employed to monitor photoexcited electrons in the two-dimensional hybrid perovskite (BA)$_2$(MA)$_2$Pb$_3$I$_{10}$. Photoelectron intensity maps are in good agreement with ab-initio calculations of the band structure. The effective mass is $-0.18 \pm 0.02 m_e$ and $0.12 \pm 0.02 m_e$ for holes and electrons, respectively. In the photoexcited state, spin-orbit splitting of the conduction band cannot be resolved. This sets the upper bound of photoinduced Rashba coupling to $\alpha_C<2.5$ eV\AA. The correlated electron-hole plasma evolves in Wannier excitons with Bohr radius of 2.8 nm, while no sign of self-trapping in small polarons is found within the investigated time window of up to 120 ps following photoexcitation.

\end{abstract}


\maketitle

Dynamical disorder refers to large fluctuations of the local atomic structure within the crystal. In the case of hybrid perovskites, such a phenomenon could generate ferroelectric domains \cite{Frost} and significantly affect the electronic states \cite{Ghosh,Nandi}. For example, dynamical disorder could favor the self-trapping of carriers in polaronic states \cite{Neukirch}. Alternatively, the local breakdown of inversion symmetry \cite{Joanna} may lift spin degeneracy at the band extrema, leading to counterrotating spin textures of electronic states with spin-momentum locking \cite{Etienne}. If experimentally verified, these effects would have important implications for optoelectronic properties of such hybrid materials. Despite the numerous reports, the subject is still a matter of debate, both experimentally and theoretically.

\begin{figure}[htp]
\includegraphics[width=\columnwidth]{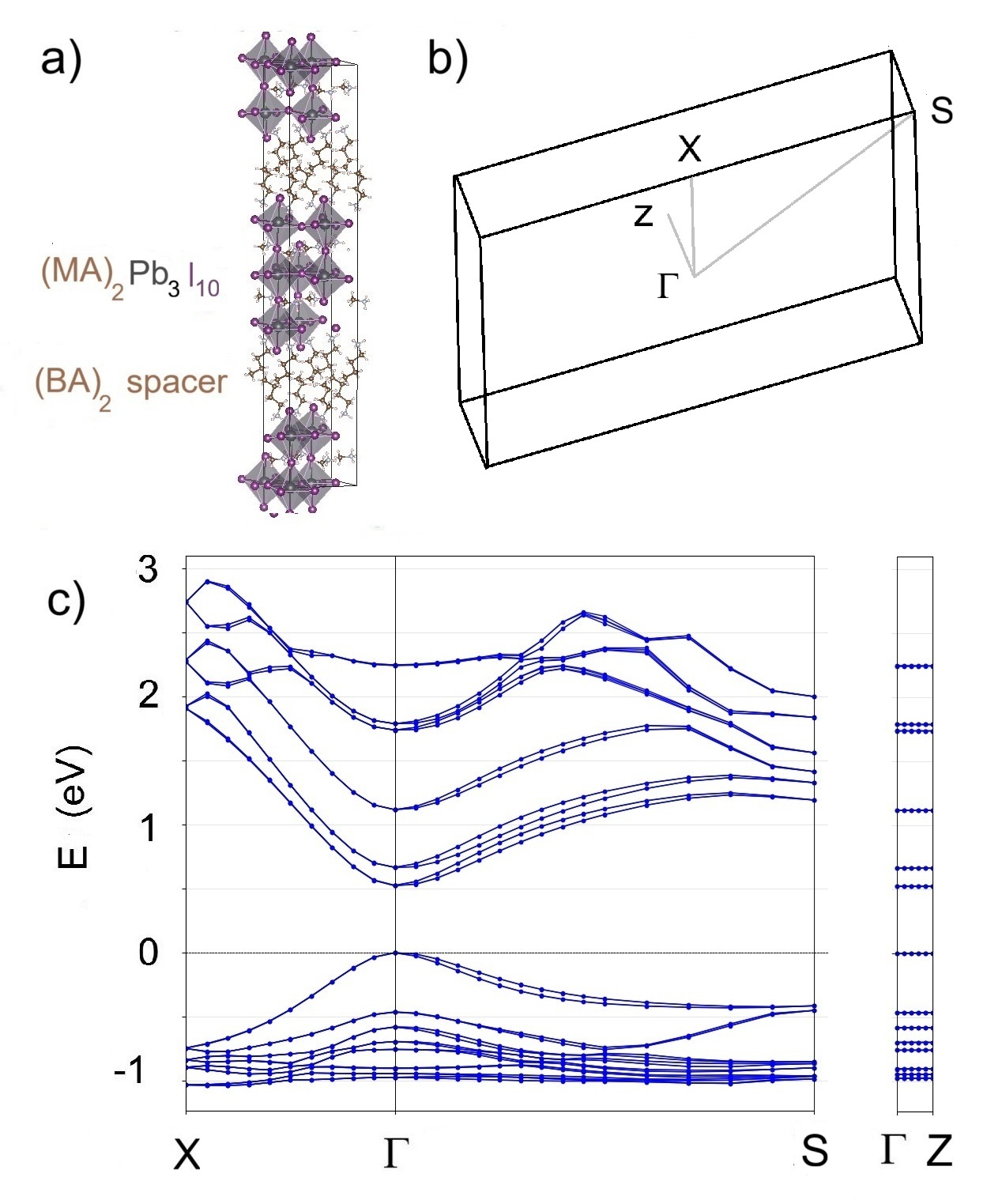}
\caption{a) Schematic structure of the 2D perovskite (BA)$_2$(MA)$_{n-1}$Pb$_n$I$_{3n+1}$ with $n=3$. b) Brillouin zone of the orthorhombic unit cell. c) Dispersion of electronic states along the $\Gamma-X$ and $\Gamma-S$ and $\Gamma-Z$symmetry directions, obtained with PBE-DFT and including spin-orbit coupling.}
\label{DFT}
\end{figure}

An ideal technique to explore the structure of electronic states out of equilibrium is time and Angle Resolved Photoelectron Spectroscopy (tr-ARPES) \cite{Ohtsubo,Dong,Chen}. In the case of a semiconductor, a visible pump laser excites electrons from the valence to the conduction band, while an ultraviolet probe pulse induces the emission of photoelectrons. An analyzer provides instantaneous snapshots of the electronic states as function of energy and wavevector \cite{Chen}. Here we employ both ARPES and tr-ARPES to map the electronic states of the two-dimensional  hybrid perovskite (BA)$_2$(MA)$_2$Pb$_3$I$_{10}$. As shown in figure \Ref{DFT}a), the structure corresponds to the $n=3$ variant of the Ruddlesden-Popper series (BA)$_2$(MA)$_{n-1}$Pb$_n$I$_{3n+1}$. It can be regarded as 3 layers of [PbI$_6$] octahedral sheets, intercalated by MA cations and sandwiched by two layers of butylammonium ligands \cite{Stoumpos}. This arrangement gives rise to multiple-quantum-well structures, in which the inorganic slabs serve as the potential well while the organic layers function as the potential barriers \cite{Liang}. Note that (BA)$_2$(MA)$_2$Pb$_3$I$_{10}$ crystallizes in the noncentrosymmetric space group $C2cb$ \cite{Stoumpos}, meaning that [PbI$_6$] octahedra reorientation develops a net dipole moment within the unit cell. Figure \Ref{DFT}b) shows the surface Brillouin zone of the orthorhombic unit cell (black lines) and of a hypothetical unit cell obtained by neglecting octahedra distortions (blue lines). The Valence Band Maximum (VBM) is located at the $\Gamma'$ point, while a weak VBM replica \cite{Park} is backfolded at $\Gamma$ by lattice distortions.

We performed PBE-DFT calculations for the experimentally detetermined structure \cite{Stoumpos} of (BA)$_2$(MA)$_2$Pb$_3$I$_{10}$. The computations employed a plane-wave energy cutoff of 40 Hartree and a 4×2×4 Monkhorst-Pack \textit{k}-point grid. Spin–orbit coupling was included via fully relativistic pseudopotentials \cite{Perdew,Setten}. Figure \Ref{DFT}c) shows the electronic band structure calculated along the high-symmetry path X(1/2,0,0) – $\Gamma$(0,0,0) – S(1/2,0,1/2) and in the out of plan direction $\Gamma(0,0,0)-Z(0,1/2,0)$. The PBE-DFT band gap at the $\Gamma$-point is found to be 0.52 eV, significantly lower than the experimental value of 2.1 eV \cite{Stoumpos}. This discrepancy is a well-known limitation of DFT calculations, which consistently underestimate band gap values. Theoretical estimate of effective masses are obtained by fitting the band dispersion along the $\Gamma - X$ path close to the band edges, resulting in $-0.16 \pm 0.01 m_0$ and $0.12 \pm 0.01 m_0$ for holes and electrons, respectively. These values are somewhat larger than reported PBEsol-DFT \cite{Stoumpos} calculations of the in-plane averaged effective masses. We have performed a PBEsol-DFT calculations and have found a $\cong 5-10$\% reduction of the hole and electron effective masses giving a good agreement with Stoumpos et al. \cite{Stoumpos} (see Fig. 5 in the supplementay information file).
The electronic bands exhibit complex Spin-Orbit splitting (SO), that varies significantly with the crystallographic direction. As shown in Fig. \Ref{DFT}c), the SO is negligible along the $\Gamma-X$ direction while attains the maximal value $\Delta E=80$ meV along the $\Gamma-S$ direction. Such anisotropic SO \cite{Blum} mainly originates from in-plane asymmetric tilting of adjacent metal halide octahedra. A small wavevector expansion along the $\Gamma-S$ direction theoretically provides a Rashba parameter $\alpha_C=0.35$ eV\AA~~for the electronic states of the conduction band. This value is consistent with the equilibrium SO splitting calculated by Jana \textit{et al.} for similar compounds \cite{Blum} and is one order of magnitude smaller than the SO splitting predicted by Etienne \textit{et al.} \cite{Etienne} for a fluctuating lattice.

\begin{figure}[htp]
\includegraphics[width=\columnwidth]{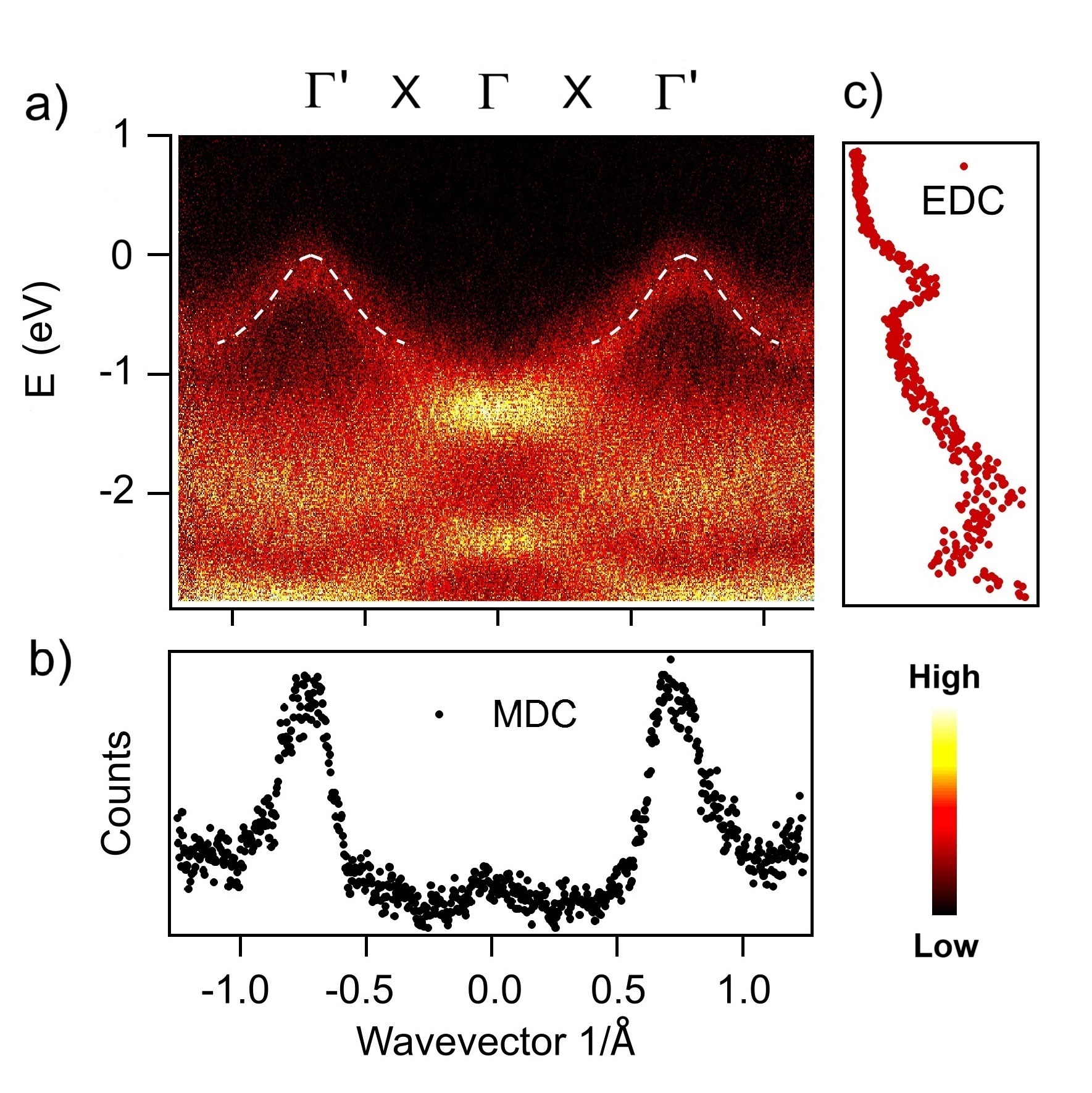}
\caption{a) Valence band dispersion acquired along the $\Gamma-X-\Gamma'$ direction. Energies have been referred with respect to the valence band maximum. The white dashed line is the valence band dispersion predicted by the \textit{ab-initio} calculations of Fig. 1c. b) Momentum Distribution Curve (MDC) extracted by integrating in an energy interval of 50 meV around the Valence Band Maximum (VBM). c) EDC extracted by integrating in a wavevector interval of 0.05 \AA$^{-1}$ around $\Gamma'$.}
\label{VB}
\end{figure}

Bulk single crystals of (BA)$_2$(MA)$_2$Pb$_3$I$_{10}$ were grown by the temperature gradient method \cite{Lin}. ARPES maps have been acquired on the TEMPO beamlime of Synchrotron Soleil, with a hemispherical analyzer MBS A-1 equipped with vacuum tunable lens axis. The photon beam is centered at 55 eV, horizontally polarized and focused down to 100 $\times$ 300 microns. Single crystals of (BA)$_2$(MA)$_2$Pb$_3$I$_{10}$ have been cleaved in ultra-high vacuum conditions and measured at 80 K. The natural cleavage plane lies in the buffer layer of the BA ligands, exposing a highly passivated surface. However two-dimensional hybrid perovskites can be easily damaged by XUV photons, so that no ARPES data have ever been reported yet. In this work, we have successfully overcome this limitation by implementing a rapid rastering acquisition mode for ARPES and tr-ARPES, enabling the direct measurement of the band structure in these materials (details are given in the supplementary information file).

As shown in Fig. \Ref{VB}a) the electronic states disperse along the $\Gamma-X-\Gamma'$ direction, reaching the VBM at $\Gamma'$. By integrating the intensity map around the VBM, we obtain a Momentum Distribution Curve (MDC) exhibiting peaks at $\Gamma'$. Also note that a weak, folded replica is generated at $\Gamma$ by octahedral distortions. Instead, by integrating the intensity map around $\Gamma'$ we obtain an Energy Distribution Curve (EDC). Of the 400 meV peak broadening, a contribution of $2\Sigma_I=100$ meV can be ascribed, via \textit{ab-initio} simulations, to the intrinsic electron-phonon linewidth \cite{ElectronPhonon}, while the remaining 300-meV contribution is likely to be dominated by inhomogeneous and extrinsic factors. 

The PBE-DFT calculations (dashed white line in Fig. \Ref{VB}a) are consistent, within the experimental spectral broadening, to the ARPES intensity map. By fitting EDCs we obtain $m_V=-0.18 \pm 0.02 m_0$. This effective mass is similar to the recently reported values in (MA)PbI$_3$ \cite{Park} or CsPbBr$_3$ \cite{Puppin,Sajedi}. We stress that $m_V=-0.18 m_0$ should be considered as a \textbf{\emph{bare}} mass value, i.e. not accounting for the renormalisation effects of electron-phonon coupling \cite{Blue}. The reason is that ARPES intensity maps vary over an energy interval that is an order of magnitude larger than the energy scale of the longitudinal optical phonons (i.e. $\hbar\Omega \cong 20$ meV) \cite{ElectronPhonon}.

\begin{figure}[htp]
\includegraphics[width=\columnwidth]{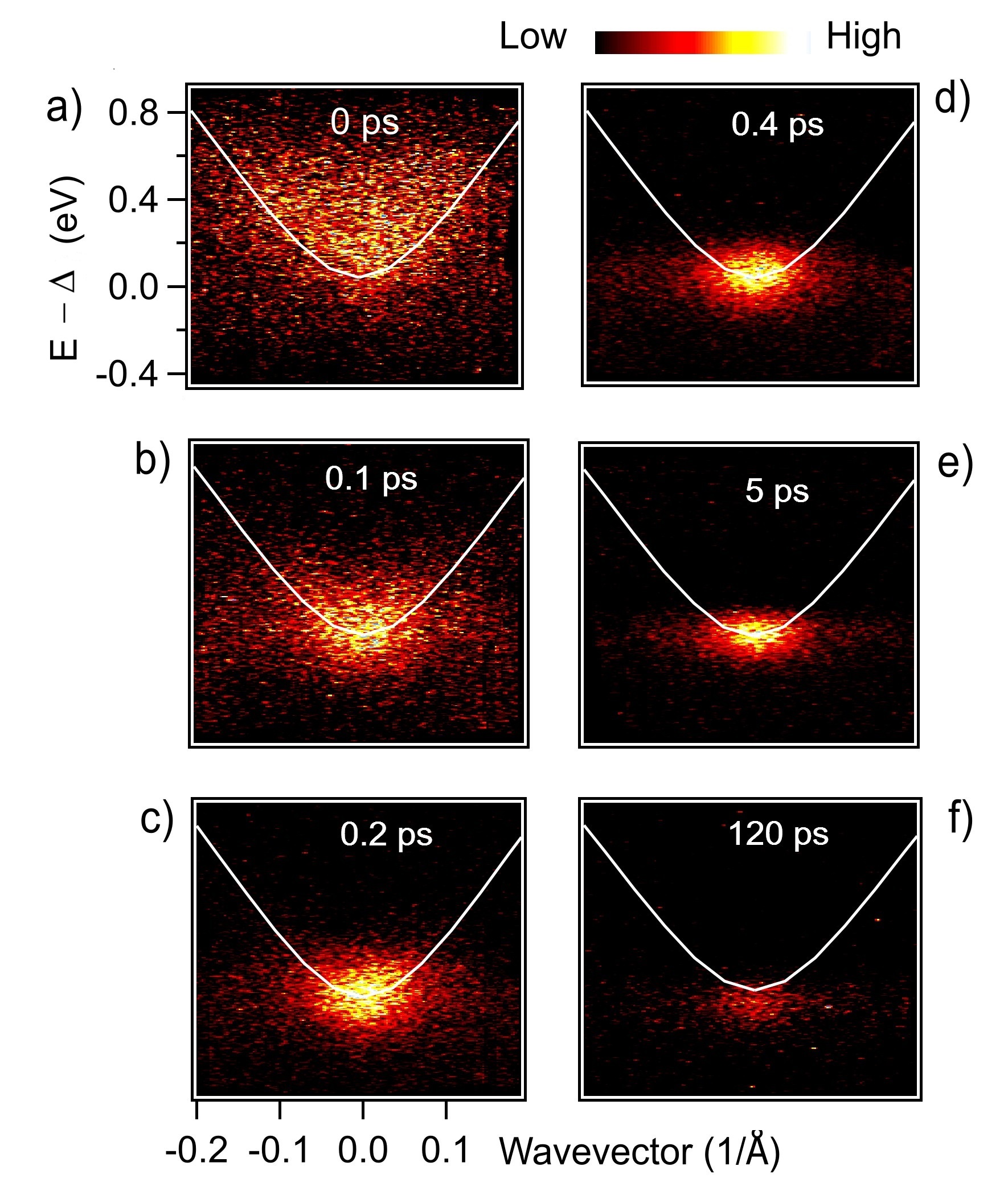}
\caption{a-f) Time resolved ARPES map along $\Gamma-X$ for different pump probe delays. With respect to the valence band map, the energy scale has been shifted by the optical gap $\Delta$. The white line is the conduction band dispersion predicted by \textit{ab-initio} calculations of Fig. 1c.}
\label{CB}
\end{figure}

\begin{figure}[htp]
\includegraphics[width=\columnwidth]{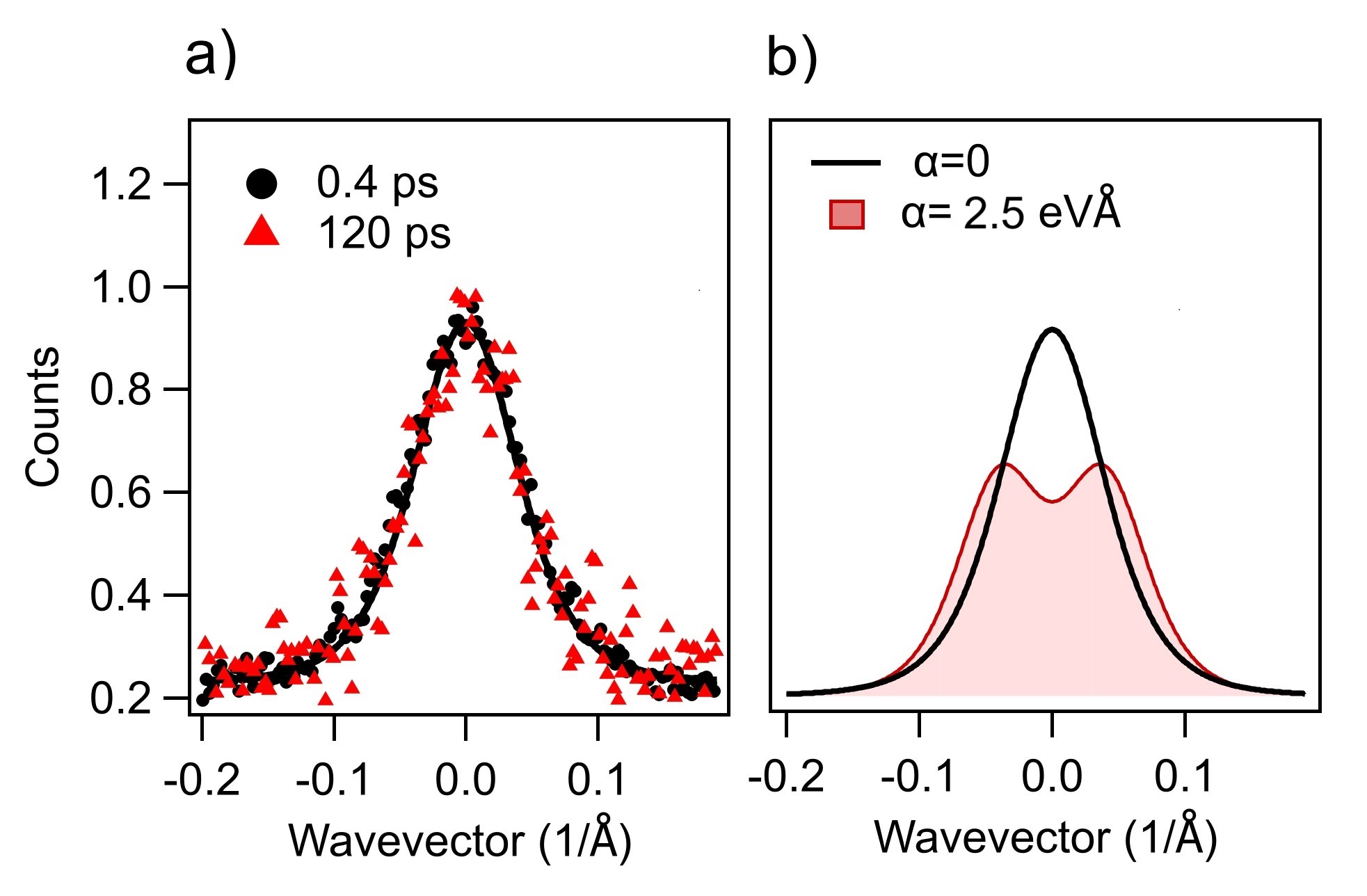}
\caption{Momentum Distribution Curves extracted at the CBM for delay times of $0.4$ ps (dark circles) and 120 ps (red triangles). The MDCs have been renormalized to the maximum value and fit (solid line) with the model function $|\phi(k)|^2$ described in the text b) Best fitting function of the MDC (black curve) and model MDC curve assuming $\alpha_C=2.5$ eV\AA~ (shadow pink area).}
\label{Rashba}
\end{figure}

The dynamics of electronic states in the conduction band is obtained via a two photon process. The probe beam, is horizontally polarized, has photon energy of $6.2$ eV and is focused down to 100 $\times$ 100 microns. The pump beam has photon energy of 3.1 eV and has incident fluence of 10 $\mu$J/cm$^2$, corresponding to photoexcitation density $\rho\cong 1\times 10^{12}$ cm$^{-2}$. The cross correlation between pump and probe has a Full Width Half Maximum of roughly 120 fs \cite{Faure}. Single crystals of (BA)$_2$(MA)$_2$Pb$_3$I$_{10}$ are cleaved at 300 K in ultra-high vacuum conditions and measured at temperature below 140 K. Also in this case, the sample has been rastered during data acquisition (see also supplementary information file).

Figure \Ref{CB}a-f shows the tr-ARPES intensity maps acquired along the $\Gamma-X$ direction for different pump-probe delays. The electrons are highly excited by the 3.1 eV pump beam and partially relax within the duration of the pump pulse. At zero delay (see Fig. \Ref{CB}a), the transient signal extends up to 0.5 eV above the Conduction Band Minimum (CBM) and it is widely distributed at energies greater than the \textit{ab-initio} calculated dispersion of the conduction band (white line in Fig. \Ref{CB}). The electron effective mass, which has never been experimentally reported in any hybrid perovskite, is determined here to be $m_C=0.12 \pm 0.2 m_0$. 

Figure \Ref{CB}b shows that hot electrons underwent strong energy relaxation as early as 0.1 ps after photoexcitation, confirming the occurrence of strong inelastic scattering. Similar findings have been previously observed in 3D hybrid perovskite (MA)PbI$_3$ \cite{Chen2, Cherasse} and in four cations mixtures \cite{Jung,Marie}. In all these cases, the energy relaxation mainly occur by collision with stretching and libration modes of the [PbI$_6$] octahedra \cite{ElectronPhonon}. The reported value of inverse quasiparticles lifetime $2\Sigma_I \cong 100$ meV \cite{ElectronPhonon} leads to cooling rate of $ 2\Sigma_I \Omega$, so that hot electrons with average excess energy $\langle E \rangle = 0.5$ eV are expected to relax in $\cong \langle E \rangle/(2\Sigma_I\Omega) \cong 0.15 $ ps \cite{Chen}. This prediction is in line with tr-ARPES maps of Figure \Ref{CB}a-c.

After 0.4 ps (see Fig. \Ref{CB}d) the electrons have relaxed, either forming a ionized carriers near to the conduction band minimum \cite{Pagliara}, or already forming hot excitons. The second hypothesis is more likely as long as the photoexcitation density is not large enough to destabilize the bound el-h states. To prove that excitons are stable at our photoexcitation density, we monitored the tr-ARPES intensity while scanning a pump beam across the optical bandgap. Figure 4 in the supplementary information file shows an exciton resonance peaking at an optical gap of 2.1 eV. The resonance signal corresponds to the photoexcitation density $\rho$, confirming the possible formation of excitons also for the above gap pumping and at early delay time (0.4 ps). As can be seen in Fig. \Ref{CB}d-f, the ARPES intensity decreases substantially (more precisely a factor 3) when moving from a delay of 0.4 ps to 120 ps. Unlike in the case of 3D perovskites \cite{Chen2}, the reduced signal intensity is entirely due to carrier recombination. The reason is that the BA ligand hinders the interlayer hopping of electrons, preventing the carrier diffusion from the surface to the bulk. Notice that the ARPES signal of Fig. \Ref{CB}d-f is sharply peaked at the center of the Brillouin Zone and shifts to lower energy by about 60 meV at delay times of 120 ps (see also Fig. 3 in supplementary information file). Such an energy shift suggests the evolution from dense and hot exciton ensemble (at 0.4 ps) to a low density exciton gas (at 120 ps).

Since our excitation is not resonant, the ARPES intensity maps in Fig. \Ref{CB}d–f do not exhibit the negative exciton dispersion that has been reported in transition metal dichalcogenides \cite{Dani,Ernstorfer}. This Floquet dressing is not always observed \cite{Dani2,Mathias,Hofer}, because is enhanced by resonant excitation conditions and may depend on specific material system. Theoretical studies also indicate that, in the case of excitation above the gap, the signal is dominated the thermal and incoherent population of excitons with finite center-of-mass momentum \cite{Knorr}.

Further conclusions can be drawn from the distribution of spectral weight in wavevector space. Figure \Ref{Rashba}a shows the MDCs extracted by integrating the ARPES intensity map in an energy interval of 0.1 eV around the CBM, at a delay time of 0.4 ps (dark circles) and 120 ps (red triangles). The two MDCs can be nicely superimposed by applying a relative multiplication factor. They both peak at $\Gamma$ and can be fitted by the function $|\phi(k)|^2=1/(1+a_0^2k^2/4)^3$ convoluted by the wavevector resolution of the setup. Here $\phi(k)$ is the 2D Fourier transform of $\phi(r)\propto \exp(-2r/a_0)$, namely the model function of a two dimensional exciton with Bohr radius $a_0$. The simple 2D hydrogenic model yields $a_0 \cong 2-4$ nm when considering the exciton effective mass $|m_C m_V|/(|m_C|+|m_V|)$, together with an effective in-plane dielectric constant $\varepsilon \cong 3-5$. The fitting of the experimental data results in $a_0=2.8$ nm, which indeed falls within the expected range. Furthermore, the exciton extension $\sqrt{\langle r^2 \rangle}=\sqrt{3/2}a_0=3.4$ nm is in good agreement with the spectroscopy of Landau levels at high magnetic fields \cite{Paulina}.

The area covered by $\phi(r)$ (i.e. $\pi\langle r ^2\rangle=36$ nm$^2$) contains 45 unit cells, therefore excluding the self-trapping of excitons in polaronic cages having the size of inter-ionic distance. This result is important for the ongoing debate about whether excitons are trapped extrinsically or intrinsically in these kinds of materials \cite{Smith}. Indeed, the formation of localized, Frenkel-like, excitons would lead to a dispersionless state across the wave-vector window considered here. We conclude that the subgap emission observed in (BA)$_2$(MA)$_{n-1}$Pb$_n$I$_{3n+1}$ \cite{Wang} are likely due to exciton trapping by a small concentration of extrinsic defects, which may be visible in integrated photo-luminescence but is irrelevant in a tr-ARPES experiment.

The last topic to discuss is dynamical Spin Orbit splitting (SO). Transient spin polarization of electronic states with $\alpha_C> 10$ eV\AA~have been proposed by theoretical simulations \cite{Etienne} and claimed by the spectral analysis of two photon absorption \cite{Lafalace}. The proposed conjecture is that ionic lattice motion and dynamical formation of fluctuating ferroelectric domains could locally increase the SO splitting. This effect could be even boosted if excited electrons would self trap in polaronic states. However, designing experiments that could ambiguously estimate the dynamical SO is especially challenging. For example, the difference in threshold energy between one-photon and two-photon absorption could be dominated by the optical selection rules of the exciton lines rather than SO splitting. The tr-ARPES is a far more direct approach for monitoring the potential presence of dynamical Rashba spitting. Figure \Ref{Rashba}b plots the fitting curve of the experimental MDCs (black line), and the MDC model assuming $\alpha_C=2.5$ eV\AA~ (pink shadow area). The absence of a double peak in the experimental MDC let us conclude that $\alpha_C<2.5$ eV\AA. Same result is obtained analyzing data acquired along the $\Gamma-S$ direction. This upper bound of $\alpha_C$ is consistent with our \textit{ab-initio} calculations and it excludes that fluctuating ferroelectric domains can generate a dynamical Rashba splitting 10 times larger than the frozen lattice value.

In conclusion, ARPES and TR-ARPES experiments on 2D hybrid Perovkistes enable us to visualise the dispersion of electronic states within the valence and conduction bands. This challenging measurement has only been possible here thanks to sample rastering during data acquisition. Our intensity maps show experimental effective masses in good agreement with first-principles calculations. First, excited electrons relax to the bottom of the conduction band, and then they form bound states with the underlying holes. The wavevector tomography of the resulting wavefunction is compatible with Wannier excitons with a Bohr radius of 2.8 nm. No evidence of self-trapping is observed until 120 ps after photoexcitation. Moreover, a thorough analysis of the momentum distribution curve also excludes that fluctuating ferroelectric domains can give rise to dynamic Rashba splitting of very large amplitude.

%First, PbI$_2$, methanamine hydriodide, butylammonium iodide, hypophosphorous acid solution and hydroiodic acid were initially formulated as the precursor solutions in 20 ml glass bottles. Then, such precursor solutions were sealed and stirred at room temperature for 30 minutes. After that, they were heated at 80 $^\circ$C until solutions become completely clear. The bulk single crystals were grown from the clear solutions at a cooling rate of 0.5 $^\circ$C/day starting from 55 $^\circ$C.

We acknowledge financial support of the 2D-HYPE project from the Agence Nationale de la Recherche (ANR, Nr. ANR 21-CE30-0059), of the Deutsche Forschungsgemeinschaft (DFG, German Research Foundation, Nr. 490867834), of the MINTOAURE project (ANR A-22-PETA-0015) and of the R\'egion Ile de France via (FemtoARPES2.0 project N. EX079194). Weiyan Qi thanks the China Scholarship Council (CSC). Computational time was granted by GENCI (Project No. 544).


\begin{thebibliography}{99}

\bibitem{Frost}
J. M. Frost, K. T. Butler, F. Brivio, C. H. Hendon, M. van Schilfgaarde, A. Walsh, Nano Lett. \textbf{14}, 2584 (2014).

\bibitem{Ghosh} D. Ghosh, E. Welch, A. J. Neukirch, A. Zakhidov, S. Tretiak, J. Phys. Chem. Lett. \textbf{11}, 3271 (2020).

\bibitem{Nandi} P. Nandi, S. Shin, H. Park, Y. In, U. Amornkitbamrung, H. J. Jeong, S. J. Kwon, H. Shin, Sol. RRL \textbf{8}, 2400364 (2024).

\bibitem{Neukirch}
A. J. Neukirch, W. Nie, J.-C. Blancon, K. Appavoo, H. Tsai, M. Y. Sfeir, C. Katan, L. Pedesseau, J. Even, J. J. Crochet, G. Gupta, A. D. Mohite, S. Tretiak, Nano Lett. \textbf{16}, 3809 (2016).

\bibitem{Joanna}
J. M. Urban, M. S. Spencer, M. Frenzel, G. Tripp\'e-Allard, M. Cherasse, C. Berrezueta-Palacios, O. Minakova, E. B. Barros, L. Perfetti, S. Reich, M. Wolf, E. Deleporte, S. F. Maehrlein, https://arxiv.org/abs/2503.02529

\bibitem{Etienne}
T. Etienne, E. Mosconi, F. De Angelis, J. Phys. Chem. Lett. \textbf{7}, 1638 (2016).

\bibitem{Ohtsubo}
Y. Ohtsubo, J. Mauchain, J. Faure, E. Papalazarou, M. Marsi, P. Le F\`evre, F. Bertran, A. Taleb-Ibrahimi, L. Perfetti, Phys. Rev. Lett. \textbf{109}, 226404 (2012).

\bibitem{Dong}
J. Dong, D. Shin, E. Pastor, T. Ritschel, L. Cario, Z. Chen, W. Qi, R. Grasset, M. Marsi, A. Taleb-Ibrahimi, N. Park, A. Rubio, L. Perfetti, E. Papalazarou, 2D Materials \textbf{10}, 045001 (2023).

\bibitem{Chen}
Z. Chen, C. Giorgetti, J. Sjakste, R. Cabouat, V. Véniard, Z. Zhang, A. Taleb-Ibrahimi, E. Papalazarou, M. Marsi, A. Shukla, J. Peretti, L. Perfetti Phys. Rev. B \textbf{97}, 241201 (2018).

\bibitem{Stoumpos}
C. C. Stoumpos, D. H. Cao, D. J. Clark, J. Young, J. M. Rondinelli, J. I. Jang, J. T. Hupp, M. G. Kanatzidis, Chem. Mater. \textbf{28}, 2852 (2016).

\bibitem{Liang}
Y. Chen, Y. Sun, J. Peng, J. Tang, K. Zheng, and Z. Liang, Adv. Mater. \textbf{30}, 1703487 (2018).

\bibitem{Park}
J. Park, S. Huh, Y. W. Choi, D. Kang, M. Kim, D. Kim, S. Park, H. J. Choi, C. Kim, Y. Yi, ACS Nano \textbf{18}, 7570 (2024).

\bibitem{Perdew}
J. P. Perdew, K. Burke, M. Ernzerhof, Phys. Rev. Lett. \textbf{77}, 3865 (1996).

\bibitem{Setten}
M. J. van Setten, M. Giantomassi, E. Bousquet, M. J. Verstraete, D.R. Hamann, X. Gonze, G.-M. Rignanese, Comput. Phys. Commun. \textbf{226} 39 (2018).

\bibitem{Blum}
M. K. Jana, R. Song, Y. Xie, R. Zhao, P. C. Sercel, V. Blum, D. B. Mitzi, Nat. Commun. \textbf{12}, 4982 (2021)

\bibitem{Lin}
W. Lin, M. Liang, Y. Niu, Z. Chen, M. Cherasse, J. Meng, X. Zou, Q. Zhao, H. Geng, E. Papalazarou, M. Marsi, L. Perfetti, S. E. Canton, K. Zheng, T. Pullerits, J. Mater. Chem. C \textbf{10}, 16751 (2022).

%\bibitem{Deng}
%S. Deng, E. Shi , L. Yuan, L. Jin, L. Dou, L. Huang, Nat. Commun. \textbf{11}, 664 (2020).

\bibitem{ElectronPhonon}
A. D. Wright, C. Verdi, R. L. Milot, G. E. Eperon, M. A. Perez-Osorio, H. J. Snaith, F. Giustino, M. B. Johnston and L. M. Herz, Nature Comm. \textbf{7}, 11755 (2016).

\bibitem{Puppin}
M. Puppin, S. Polishchuk, N. Colonna, A. Crepaldi, D. N. Dirin, O. Nazarenko, R. De Gennaro, G. Gatti, S. Roth, T. Barillot, L. Poletto, R. P. Xian, L. Rettig, M. Wolf, R. Ernstorfer, M. V. Kovalenko, N. Marzari, M. Grioni, and M. Chergui, Phys. Rev. Lett. \textbf{124}, 206402 (2020).

\bibitem{Sajedi}
M. Sajedi, M. Krivenkov, D. Marchenko, J. S\'anchez-Barriga, A. K. Chandran, A. Varykhalov, E. D. L. Rienks, I. Aguilera, S. Bl\"ugel, Oliver Rader, Phys. Rev. Lett. \textbf{128}, 176405 (2022).

\bibitem{Blue}
L. Perfetti, S. Mitrovic, G. Margaritondo, M. Grioni, L. Forr\'o, L. Degiorgi, H. H\"ochst, Phys. Rev. B \textbf{66},075107 (2002).

%\bibitem{Sajedi1} 
%M. Sajedi, M. Krivenkov, D. Marchenko, E. D. L. Rienks, A. Varykhalov, and O. Rader, Phys. Rev. B \textbf{102}, 081116(R) (2020).

\bibitem{Faure}
 J. Faure, J. Mauchain, E. Papalazarou, W. Yan, J. Pinon, M. Marsi, L. Perfetti, Rev. Sci. Instrum. \textbf{83}, 043109 (2012).

%\bibitem{Chen1}
%Z. Chen, J. Sjakste, J. Dong, A. Taleb-Ibrahimi, J.-P. Rueff, A. Shukla, J. Peretti, E. Papalazarou, M. Marsi, and L. Perfetti, PNAS \textbf{36}, 21962 (2020).

\bibitem{Chen2}
Z. Chen, M.-I. Lee, Z. Zhang, H. Diab, D. Garrot, F. Lédée, P. Fertey, E. Papalazarou, M. Marsi, C. Ponseca, E. Deleporte, A. Tejeda, L. Perfetti, Phys. Rev. Materials \textbf{1}, 045402 (2017).

\bibitem{Cherasse}
M. Cherasse, J. Dong, G. Tripp\'e-Allard, E. Deleporte, D. Garrot, S. F. Maehrlein, M. Wolf, Z. Chen, E. Papalazarou, M. Marsi, J.-P. Rueff, A. Taleb-Ibrahimi, L. Perfetti, Nano Letters \textbf{22}, 2065 (2022).

\bibitem{Pagliara}
V. Gosetti, J. Cervantes-Villanueva, S. Mor, D. Sangalli, A. Garcia-Cristobal, A. Molina-Sanchez, V. F. Agekyan, M. Tuniz, D. Puntel, W. Bronsch, F. Cilento, and S. Pagliara, Progress in Surface Science \textbf{100}, 100777 (2025).

\bibitem{Jung}
E. Jung, K. Budzinauskas, S. Oz, F. Unlu, H. Kuhn, J. Wagner, D. Grabowski, B. Klingebiel, M. Cherasse, J. Dong, P. Aversa, P. Vivo, T. Kirchartz, T. Miyasaka, P. H. M. van Loosdrecht, L. Perfetti, S. Mathur, ACS Energy Letters \textbf{5}, 785 (2020).

\bibitem{Marie}
M. Cherasse, N. Heshmati, J. M. Urban, F. Unlu, M. S. Spencer, M. Frenzel, L. Perfetti, S. Mathur, S. F. Maehrlein, Small \textbf{21}, 2500977 (2025).

\bibitem{Dani}
M. K. L. Man, J. Mad\'eo, C. Sahoo, K. Xie, M. Campbell, V. Pareek, A. Karmakar, E. L. Wong, A. Al-Mahboob, N. S. Chan, D. R. Bacon, X. Zhu, M. M. M. Abdelrasoul, X. Li, T. F. Heinz, F. H. da Jornada, T. Cao, K. M. Dani, Science Advances \textbf{7}, eabg0192 (2021).

\bibitem{Ernstorfer}
S. Dong, M. Puppin, T. Pincelli, S. Beaulieu, D. Christiansen, H. H\"ubener, C. W. Nicholson, R. P. Xian, M. Dendzik, Y. Deng, Y. W. Windsor, M. Selig, E. Malic, A. Rubio, A. Knorr, M. Wolf, L. Rettig, R. Ernstorfer, Natural Sciences \textbf{1}, e10010 (2021).

\bibitem{Dani2}
J. Mad\'eo, M. K. L. Man, C. Sahoo, M. Campbell, V. Pareek, E. L. Wong, A. Al-Mahboob, N. S. Chan, A. Karmakar, B. M. K. Mariserla, X. Li, T. F. Heinz, T. Cao, K. M. Dani, Scince \textbf{370} 1199, (2020).

\bibitem{Mathias}
D. Schmitt, J. P. Bange, W. Bennecke, A. AlMutairi, G. Meneghini, K. Watanabe, T. Taniguchi, D. Steil, D. R. Luke, R. T. Weitz, S. Steil, G.S. Jansen, S. Brem, E. Malic, S. Hofmann, M. Reutzel, S. Mathias, Nature \textbf{608}, 499 (2022).

\bibitem{Hofer}
R. Wallauer, R. Perea-Causin, L. M\"unster, S. Zajusch, S. Brem, J. G\"udde, K. Tanimura, K.-Q. Lin, R. Huber, E. Malic, U. H\"ofer, Nano Lett. \textbf{21} 5867 (2021).

\bibitem{Knorr}
D. Christiansen, M. Selig, E. Malic, R. Ernstorfer, and A. Knorr, Phys. Rev. B 100 205401 (2019).

\bibitem{Paulina}
M. Dyksik, S. Wang, W. Paritmongkol, D. K. Maude, W. A. Tisdale, M. Baranowski, and P. Plochocka, J. Phys. Chem. Lett. \textbf{12}, 1638 (2021).

%\bibitem{Deng}
%S. Deng, E. Shi, L. Yuan, L. Jin, L. Dou, L. Huang, Nature Comm. \textbf{11}, 664 (2020).

\bibitem{Smith}
M. D. Smith, H. I. Karunadasa, Acc. Chem. Res. \textsl{51}, 619 (2018).

%\bibitem{Kong}
%J. Yu, J. Kong, W. Hao, X. Guo, H. He, W. R. Leow, Z. Liu, P. Cai, G. Qian, S. Li, X. Chen, X. Chen, Advanced Materials \textbf{31}, 1806385 (2019).

\bibitem{Wang}
J. Li, J. Wang, J. Ma, H. Shen, L. Li, X. Duan, D. Li, Nature Comm. \textbf{10}, 806 (2019).

\bibitem{Lafalace}
E. Lafalce, E. Amerling, Z.-G. Yu, P. C. Sercel, L. Whittaker-Brooks, Z. V. Vardeny, Nature Comm. \textbf{13}, 483 (2022).

%\bibitem{Giustino}
%M. Schlipf, F. Giustino, Phys. Rev. Lett. \textbf{127}, 237601 (2021).

\end{thebibliography}
\end{document}